\documentclass[%
 reprint,
 amsmath,amssymb,
 aps,
floatfix,
]{revtex4-1}

\usepackage{caption}
\usepackage{subcaption}
\usepackage{mathtools}
\usepackage{pgfplots}
\pgfplotsset{compat=1.17}
\usepackage{upgreek}

\begin{document}

\title{Emergence of Accurate Atomic Energies from Machine Learned Noble Gas Potentials}

\author{Frank Uhlig}\thanks{Contributed equally}
 \affiliation{Institute for Computational Physics, University of Stuttgart, Stuttgart, Germany}
\author{Samuel Tovey\textsuperscript{*}}%
 \affiliation{Institute for Computational Physics, University of Stuttgart, Stuttgart, Germany}
\author{Christian Holm}
 \email{holm@icp.uni-stuttgart.de}
\affiliation{Institute for Computational Physics, University of Stuttgart, Stuttgart, Germany}

\date{\today}

\begin{abstract}
The quantum theory of atoms in molecules (QTAIM) gives access to well-defined local atomic energies.  
Due to their locality, these energies are potentially interesting in fitting atomistic machine learning models as they inform about physically relevant properties. 
However, computationally, quantum-mechanically accurate local energies are notoriously difficult to obtain for large systems.  
Here, we show that by employing semi-empirical correlations between different components of the total energy, we can obtain well-defined local energies at a moderate cost.
We employ this methodology to investigate energetics in noble liquids or argon, krypton, and their mixture.
Instead of using these local energies to fit atomistic models, we show how well these local energies are reproduced by machine-learned models trained on the total energies.
The results of our investigation suggest that smaller neural networks, trained only on the total energy of an atomistic system, are more likely to reproduce the underlying local energy partitioning faithfully than larger networks.
Furthermore, we demonstrate that networks more capable of this energy decomposition are, in turn, capable of transferring to previously unseen systems.
Our results are a step towards understanding how much physics can be learned by neural networks and where this can be applied, particularly how a better understanding of physics aids in the transferability of these neural networks.
\end{abstract}

\maketitle
\paragraph*{Introduction}
Machine learning (ML) has impacted many aspects of modern society.
From artificial intelligence, smart homes, and social media to an industrial context, e.g., predictive maintenance or medical diagnosis, ML methods have changed how we interact with computers and our environment.  
In recent years, ML methods have been exploited in the development of inter-atomic potentials for liquids~\cite{tovey20a, daru22a}, solids~\cite{sivaraman20a, shishvan23a, liang23a} and interfaces~\cite{rowe21a, molpeceres23a} with almost ab initio accuracy but at a highly reduced cost for running associated molecular dynamics (MD) simulations.
Many reviews have been dedicated to ML and its various applications to potential energy surfaces~\cite{handley2010}, computational/theoretical chemistry~\cite{goh17a, rupp18a}, and cheminformatics~\cite{mitchell2014, lo2018}.
ML can interpolate between training data without knowledge of the differential equations describing the underlying physics, thus offering considerable flexibility at the cost of interpretability in the final model.
The most common approaches to developing machine-learned potentials include directly fitting the total energy of a system~\cite{handley2010}, splitting the total energy into local atomic energies~\cite{behler07a, bartok10a}, or using ML to produce parameters for physically motivated potentials from which the energy of a system can be calculated~\cite{popelier2017, bereau2018}.  
Splitting the total energy into a sum of local atomic energies has important features: it offers accuracy, flexibility, and scalability if models are fitted only for the involved atomic species.  
This is achieved by representing atomic environments centred on individual atom sites as feature vectors.  
Different variations of these feature vectors, commonly referred to as \emph{descriptors}, have been developed with varying success and accuracy~\cite{behler07a, bartok10a, behler11a, behler12a, rupp12a, bartok13a, thompson15a, hansen2015, lilienfeld2015, jenke2018, behnam22a, zaverkin20a}.
However, it is unclear whether assuming that a system's total energy can be split into a simple sum of atomic energies is valid.  
It can be argued that, due to the locality of quantum mechanics, the energy of an atom is mainly dependent on its local environment~\cite{thompson15a}.
Indeed, the \emph{quantum theory of atoms in molecules} (QTAIM) allows us to dissect the total quantum-mechanical energy of any system into well-defined atomic energies at stationary states of the atomic configuration~\cite{bader85a, bader1994}. 
Further development has enabled the calculation of atomic energies for non-stationary states by directly calculating them from density matrices of the system.~\cite{popelier2001} 
This has been exploited to train machine learning models to predict these local atomic energies.~\cite{popelier2015, popelier2016} 
One can then accurately predict atomic energies and effective atomic charges\cite{popelier2017} that can be used to describe the interaction of atoms and molecules. 
One of the most challenging aspects of machine learning is to provide transferable models that can be trained on a small set of reference data and applied to a more extensive set with comparable accuracy~\cite{batatia23a}.
In the field of computer vision, some transferability is assumed in the context of transfer learning, i.e., refining pre-trained models for a specific task.
Similarly, models for mixed component systems have been systematically fitted on the pure components first and only then refined for the full systems~\cite{hajinazar2017}.
Recent efforts regarding the prediction electron densities show that transferability can be achieved by using atom-centered symmetry functions as descriptors~\cite{grisafi2018}.
To achieve transferability and scalability, it is crucial that the employed descriptor can suitably accommodate the inclusion of different atomic species without growing in a computationally intractable manner. 
Models trained on large subsets of with vast chemical diversity offer good performance for various tasks, from predicting energetics to identifying chemical concepts.~\cite{schuett17a} 
Current developments focus on generating transferable descriptors that offer a reduction in scaling with respect to the number of involved chemical species, and thus, training data~\cite{gastegger2018, artrith2018, unke18a, jindal2018, bereau2018, tamura18a}.
Another approach to enforce transferability is to reproduce the underlying physics more accurately.  
This could mean ensuring that the predicted local energies correspond to a physical theory in the above context. 
Here, to the best of our knowledge, for the first time, we report a direct comparison between physically meaningful local energies and energies obtained with different machine-learning algorithms for bulk liquids.  
This work investigates under which conditions machine learning models reproduce the underlying energetics and how much data is needed to learn the corresponding physics.
We argue that smaller networks are forced to learn more fundamental representations of the underlying data they fit, sometimes leading to the emergence of learned physics.

\paragraph*{Theory}
The quantum mechanical virial theorem establishes a connection between the Laplacian of the electron density \(\rho({\bf r})\), the kinetic energy density \(\mathcal{T}({\bf r})\), and the total virial \(\mathcal{V({\bf r})}\)~\cite{bader85a}:
\begin{equation}\label{eqn:virial}
\left( \frac{\hbar^2}{4m} \right) \nabla^2 \rho({\bf r}) = 2 \mathcal{T}({\bf r}) + \mathcal{V}({\bf r}).
\end{equation}
If integrated over the whole system, e.g., an isolated molecule, the left-hand side of~eq.~\ref{eqn:virial} disappears and relates the kinetic energy to the total virial.
The same holds true for zero-flux surfaces \(S({\bf r})\) of the electron density, defined by means of Gauss' divergence theorem using the gradient of the electron density and the normal vector, $\bf{n}(\bf r)$ as
\begin{equation}\label{eqn:dens}
\nabla \rho({\bf r}) \cdot \bf{n}({\bf r}) = 0
\end{equation}
which, within the framework of QTAIM, determines atomic regions, \(\Omega\).
Hence, for QTAIM, we can also define a relationship between the local virial and the kinetic energy. 
At a stationary state with respect to the nuclei, the virial becomes the local potential energy of the $i^{\text{th}}$ atom:
\begin{equation}\label{eqn:state}
-2 \mathcal{T}(\Omega_{i}) = \mathcal{ V }(\Omega_{i}) \overset{\text{stat. state}}= V(\Omega_{i})
\end{equation}
and the total energy, $E$ can be obtained as a sum of local energies of atoms in molecules by:
\begin{equation}\label{eqn:total_aim_energy}
E = \sum_{i} V(\Omega_{i}).
\end{equation}
With~eq.~\ref{eqn:total_aim_energy}, or modified versions of it~\cite{popelier2001}, one has access to well-defined local atomic energies that can be used to fit machine-learned models.~\cite{popelier2015, popelier2016}
A disadvantage of~eq.~\ref{eqn:total_aim_energy} is that the connection between potential and kinetic energy only exists as such for stationary states. 
Another caveat is that the kinetic energy density is difficult to obtain at an adequate resolution for large-scale analysis. 
Due to the involved gradient of the wavefunction, very fine grids must be used, which quickly become impractical to handle.
However, one might expect that certain correlations between energy components still exist, even for structures sampled at finite temperature.
If so, it would unlock new methods for validating and even training machine learned interatomic potentials.

In typical machine-learning approaches for the fitting of machine-learned inter-atomic potentials, a model is fit to reproduce the same total total energy of the system by first computing the local energies of so-called atomic environments, or descriptors, $\vec{g}$, and then summing over them by
\begin{equation}
    \centering
    E = \sum\limits_{i}f_{\theta}(\vec{g}_{i}),
\end{equation}
where $f_{\theta}$ is a neural network parameterised by the set $\theta$.
The parameters used in constructing these descriptors will define the environment in which the local energy is computed.
While many descriptors exist, this work uses the invariant 4D-Bispectrum descriptor in all model training~\cite{bartok13a}.
These descriptors are fixed, meaning that they do not evolve during training, and they include a species scaling to differentiate different atom types.
The network is updated via gradient descent using a loss function of the form
\begin{equation}
    \centering
    \mathcal{L} = \left(E - \sum\limits_{i}f_{\theta}(\vec{g}_{i})\right)^{2}.
\end{equation}
Under this construction, there is much ambiguity about how atomic energies are assigned to specific atoms, i.e., which partitioning scheme is learned by the network.
We argue that the correct partitioning of energies into their atomic contributions will result in better models as the underlying physics is better represented.
This can be considered a feature learning problem, as the networks must learn how to correctly represent specific atomic environments in the latent space to be transferable to unseen local environments, something only possible if the partitioning is done physically.

Feature learning, or representation learning as it is often referred to, describes the ability of a neural network to learn a fundamental representation of a data manifold rather than simply fitting to training points~\cite{bengio13a}.
For neural networks, this means that features in the data, for example certain shapes in the case of images, have been learned during propagation through layers.
In the context of fitting machine-learned inter-atomic potentials, feature learning can refer to both the learning of a descriptor internally within the neural network or, as is discussed here, learning the most physical distribution of local energies, that is, the distribution closest to the physically motivated QTAIM method.
It has been shown that feature learning is fundamental to the success of both pre-training and transfer learning paradigms~\cite{erhan10a, day17a}.
However, not all neural networks are capable of feature learning.
In their 2021 report, \textit{Yang and Hu} demonstrated that large neural networks' ability to perform feature learning strongly depends on their initialisation. 
Often, they are incapable of feature learning, becoming purely regressive machines; that is to say, they learn purely the data they are trained on and not features connecting the points~\cite{yang21a}.

It is not difficult to argue that learning a fundamental physical principle would amount to feature learning for a neural network or, to say it differently, a neural network with knowledge of physics has learned the correct features to describe a problem.
This point is often leveraged in developing physics-informed networks, which have been shown to lead to better transferrability~\cite{cuomo22a, choudhary20a, patel22a}.
However, it is not well studied whether a neural network, constrained in its size and, therefore, in the complexity of features that can be learned, will, in turn, develop knowledge of physics.
This work examines this question by determining under what conditions a neural network can learn the correct energy decomposition in an atomistic system.

\paragraph*{Local energy computation}
Before identifying the role of architecture and the emergence of correct energy partitioning, we demonstrate that it is possible to apply the QTAIM method to the extraction of local energies in the non-stationary states produced in our noble-liquid DFT simulations.
To do so, a correlation must be identified between the classical electrostatic contribution to the system's total energy and the other Hamiltonian components.
Such a correlation would allow us extract local energy information on much larger grids, thereby making the computation accessible to large, periodic systems.
To investigate if this correlation exists, we performed ab initio molecular dynamics simulations (AIMD) based on density functional theory (DFT) for three different bulk liquids, Ar(108), Kr(108), and Kr(54)Ar(54) at 120 K.
AIMD simulations used Kohn-Sham density functional theory (KS-DFT) and were carried out with \textsc{CP2K}~\cite{vandevondele05b, hutter14a}. 
The exchange-correlation energy was approximated using a van-der-Waals density functional~\cite{vydrov10a, sabatini13a}. 
The Kohn-Sham one-particle wavefunctions were expanded into an atom-centered double-\(\zeta\) basis set optimized for dense liquids\cite{vandevondele07a}. 
The electron density was expanded in an auxiliary plane-wave basis set with a kinetic energy cutoff of 600 Ry.
Initial configurations were obtained from molecular dynamics simulations using suitable noble liquid potentials. 
Newton's equations of motion were integrated using a timestep of 2 fs, and the temperature of the system was kept constant at 120 K using a Nosé-Hoover chain thermostat~\cite{martyna92a} (chain length: 3, time constant: 100 fs).
This work focuses on noble liquid mixtures as a proxy for more complicated liquids, specifically liquid argon, krypton and their equimolar mixture. 
\begin{figure}[t]
    \centering
    \includegraphics[width=\linewidth]{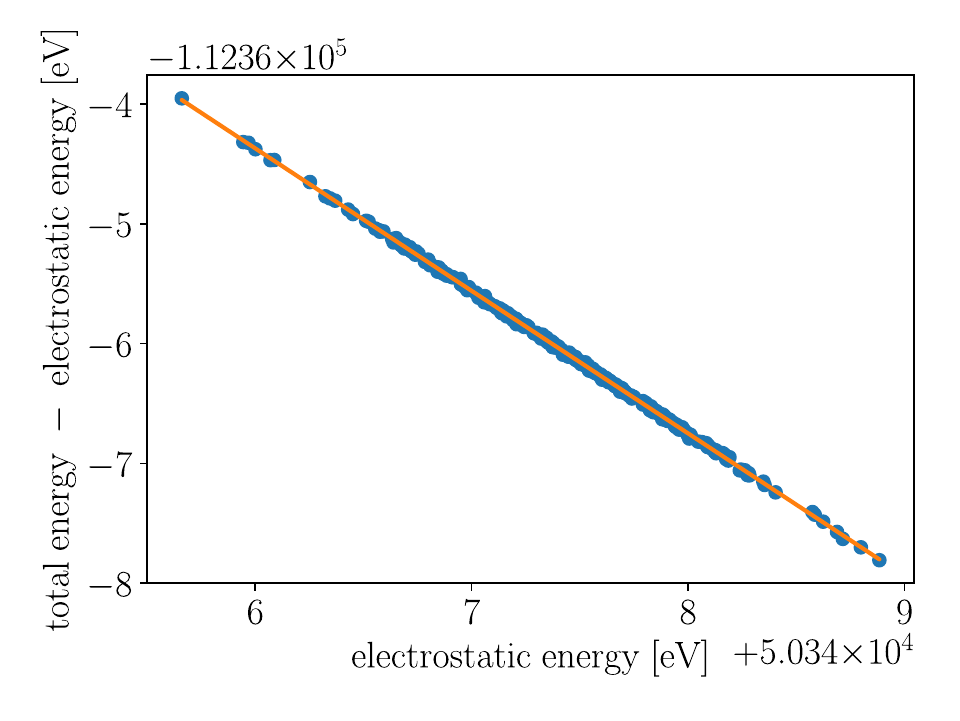}
    \caption{Total energy minus classical electrostatic energy versus the
  classical electrostatic energy in pure Ar.}
\label{fig:correlation}
\end{figure}
Figure~\ref{fig:correlation} outlines the results of this study for the liquid argon (see Figures~\ref{fig:mix-correlation} and~\ref{fig:kr-correlation} in SI for the mixture and pure Kr respectively.).
In these systems, we observe a strong correlation between the classical electrostatic energy and the sum of all other contributions to the total energy, i.e., kinetic energy and exchange-correlation energy. 
The correlation between these variables can be fit linearly, and while the values for slope and y-intercept are system-specific, they can be attributed to physical interpretations.
For the pure systems, the intercepts can be related to the classical electrostatic energy of the isolated atoms in vacuo.  
This is not a general characteristic as it does not apply to the mixed system.  
Nevertheless, this study demonstrates a way to relate the electrostatic energy effectively to the remaining energy components.  
The fit's root mean squared error is approximately 0.3 meV/atom, the same order as a typically achieved fit accuracy in machine learned potentials.
We can largely ignore this source of uncertainty as the error in the machine learning model fit will outweigh it.
Using these results, we can dissect the classical electrostatic interaction between all charged particles (electron density and protons) as it is readily available and accurately reproduced on grids of moderate resolution and can therefore be used for large, 3D-periodic systems.
The dissection follows in accordance with~eq.~\ref{eqn:state} and is considered a robust numerical procedure\cite{yu2011}.  
The electrostatic energy of an atom \(E_\Omega\) is calculated using the electron density \(\rho\) and the Hartree potential \(V_\mathrm{H}\) as \(E_\Omega = \frac{1}{2} \int_\Omega \rho V_\mathrm{H}\).  
We will refer to these energies as local DFT energies, \(E_\mathrm{loc}^\mathrm{DFT}\).

\paragraph*{Local energy fitting}
With the ability to produce reference data for the local energies, we can turn our attention to understanding under which conditions neural networks, trained purely on the total energy of ab initio simulations, can accurately reproduce the atomic energies computed using QTAIM.
To do so, neural networks of various architectures were trained on pure argon and krypton systems and the ArKr mixture using the total energy of each configuration in the loss function.
Due to the similarity of the results, only the results from a single architecture are shown in the manuscript, namely, a single layer network with hyperbolic tangent activation.
Other architectures are displayed in the appendix.
Neural networks utilised hyperbolic-tangent and linear activation functions and were trained using a mean-squared-error loss function.
In all cases, the correlation score for the total energy, the quantity on which the network was trained, remains high across the architecture space (see appendix Figure~\ref{fig:si-all-archs}).
For small models, the initialisation variance results in large error bars, which are saturated for the wider and deeper models.

\begin{figure}[t]
    \centering
        \input{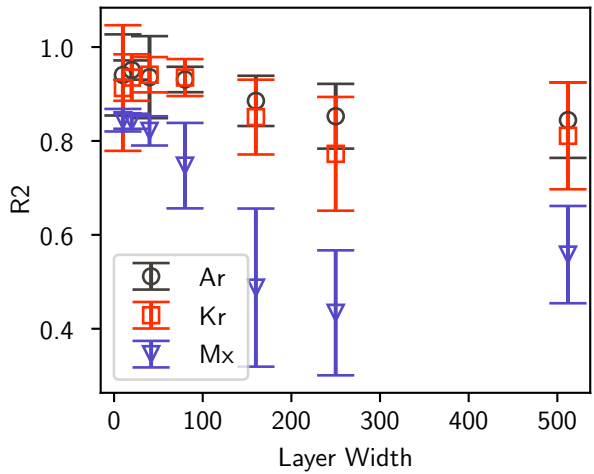}
    \caption{Correlation score for the local energy of each system, the argon (Ar), krypton (Kr), and the 50-50 mixture (Mx).}
    \label{fig:local_energy}
\end{figure}

As all model architecture are capable of fitting the total energies, the interest now lies in seeing how the change in architecture will impact the ability of the models to reproduce the local QTAIM energies of the systems correctly.
To do so, Pearson correlation coefficients are computed on the local energy test data produced from the DFT simulations.
This is data that the neural networks have never seen during their training
Figure~\ref{fig:local_energy} displays these correlation coefficients between the predicted and DFT-produced local energies for the pure and mixed systems (see appendix Figure~\ref{fig:si-local_energy} for all architectures).
Studying these local correlation plots, it is clear that larger neural networks assign the local energies more arbitrarily and show little correlation with the underlying physics.
On the other hand, smaller networks maintain a reasonable correlation between the predicted and actual partitioning of the local energy values.
It is clear, however, that the models trained on the mixtures struggle more to achieve this local energy decomposition, with these networks achieving lower maximum scores and dropping off faster than their pure system counterparts.

\begin{figure}[t]
    \centering
        \input{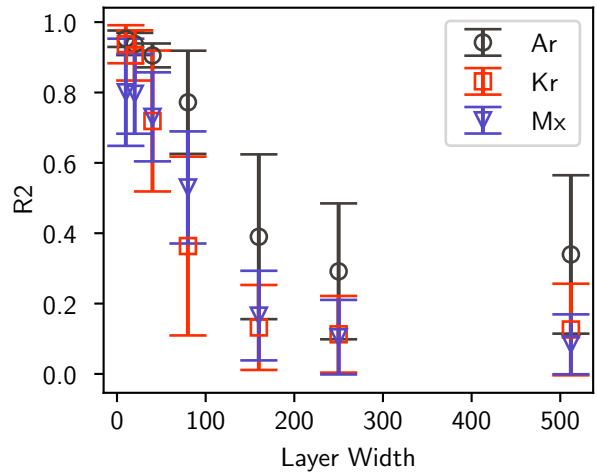}
    \caption{Correlation score for the total energy of each system computed using the transferred models, the argon (Ar), krypton (Kr), and the 50-50 mixture (Mx).}
    \label{fig:total_transfer}
\end{figure}

\begin{figure}[htb]
    \centering
        \input{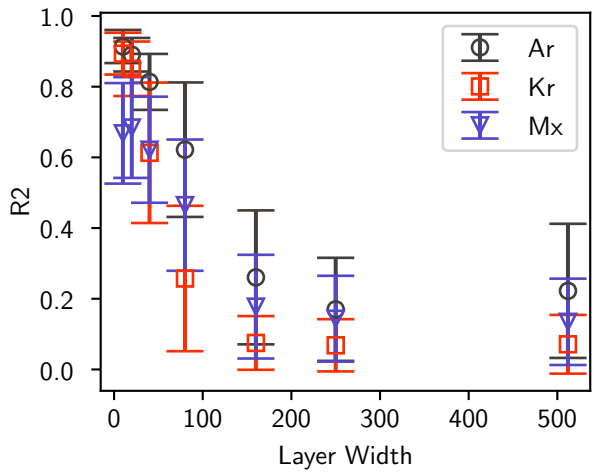}
    \caption{Correlation score for the local energy of each system computed using the transferred models, the argon (Ar), krypton (Kr), and the 50-50 mixture (Mx).}
    \label{fig:local_transfer}
\end{figure}

\paragraph*{Transferability}
Given the emergence of correct local energies, it is of interest to see if this emergent structure learning can be used in transfer studies.
We investigate this by testing if models more capable of local energy decomposition are also capable of being transferred, without re-training, to previously unseen configurations.
The models trained on the ArKr mixture are used to compute the total and local energies of the pure Ar and Kr systems whereas the models trained on the pure systems are used to compute the energies in the mixed systems.
This is done on the same test data as in the previous section using the \texttt{pair style} command in the LAMMPS simulation engine where atomic environments centered on an Ar atom will use the pure Ar model for the force computation and the pure Kr models handle the Kr centered environments.
Figure~\ref{fig:total_transfer} outlines the results of this study on the total energy computations for all architectures and systems.
As is to be expected, the models fail in their predictions early in all cases.
It is, however, interesting that the smaller models can retain a high correlation score for the transferred systems.
This suggests that these models represent the underlying physics of the problem better.
Another point of note is the ability of the Tanh activation function models to achieve more significant correlation scores for slightly longer than the linear models.
Moving away from the total energies, Figure~\ref{fig:local_transfer} shows the results of the transfer experiment on the local energy predictions.
These plots closely resemble those of the total energies in the gradual degradation of performance with increasing network width.
In typical machine learning theory, the networks addressed here would not be considered purely regressive as they are not approaching an infinite width limit.
However, even amongst the feature learning limit, there appears to be a discrepancy between architectures.
We argue that this comes down to the number of degrees of freedom each of these neural networks has and how they can use these to distribute information.
Namely, the networks that are required to find solutions with less degrees of freedom typically learn more physical representations.
Such an outcome is supported by the data and resembles the principle of Occam's razor: the simplest solution is often the correct one.

\paragraph*{Conclusion}
The decomposition of total energy into atomic contributions has been significantly utilised in machine-learned inter-atomic potentials.
However, whether these local energies can be considered physically meaningful requires clarification.
In this investigation, we have restructured this problem as one of physics-aware feature learning in neural networks.
By considering local energies obtained from the QTAIM as ground truth, we have argued that correct feature learning of a neural network involves learning these physically derived local energies given the system's total energy.
To this end, studies of pure liquid argon and krypton and their mixture have been performed wherein DFT-MD was performed to generate ab initio data.
We have shown that QTAIM can be applied to bulk liquid systems and computed these local energies.
Neural network models were then fitted to the total energies of the ab initio simulations for different architectures before the local decomposition was compared with theoretical values.
We find that neural networks of smaller width (12-100 nodes) display improved decomposition over larger models.
These results align with those of feature learning in the field of learning theory, wherein it is understood that smaller neural networks are capable of so-called feature learning, i.e., extraction of features in data, while larger models are purely regressive.
We further highlight that even among feature learning networks, those with fewer degrees of freedom learn more physically relevant representations, thus resulting in more physically correct energies.
We conclude that smaller networks are faster to train and more easily deployed whilst retaining the accuracy of larger networks. 
They also learn a more physical representation of atomic decomposition and are thus favorable from a physics perspective.
Finally, we have shown that the models capable of computing accurate local energies also showed the ability to accurately transfer to previously unseen configurations.
While it is expected that this transferability is predominantly a geometric one, future work should identify the limits to which this can be pushed in chemical space.

\begin{acknowledgments}
The authors acknowledge financial support from the German Funding Agency (Deutsche Forschungsgemeinschaft DFG) under Germany's Excellence Strategy EXC 2075-390740016.
This work was supported by SPP 2363- "Utilization and Development of Machine Learning for Molecular Applications – Molecular Machine Learning."
Funded by the Deutsche Forschungsgemeinschaft (DFG, German Research Foundation), Project-No 497249646.
S.T would like to thank David Tovey for his thorough review of and comments on the manuscript.
The authors would like to thank Anand Narayanan Krishnamoorthy and Matthias Bauer for their helpful input on early versions of this study involving Gaussian process-based models.
All authors would like to acknowledge and thank David Beyer for his detailed reading and editing of the manuscript. 
\end{acknowledgments}

\bibliography{references}
\appendix
\section{Appendix A: Energy Decomposition}
Here we show the energy decomposition for the pure krypton and argon-krypton mixture systems.
These results highlight that using the electrostatic component of the energy to infer local energies is valid across all studied systems.
\begin{figure*}[t]
\centering
\includegraphics[width=\textwidth]{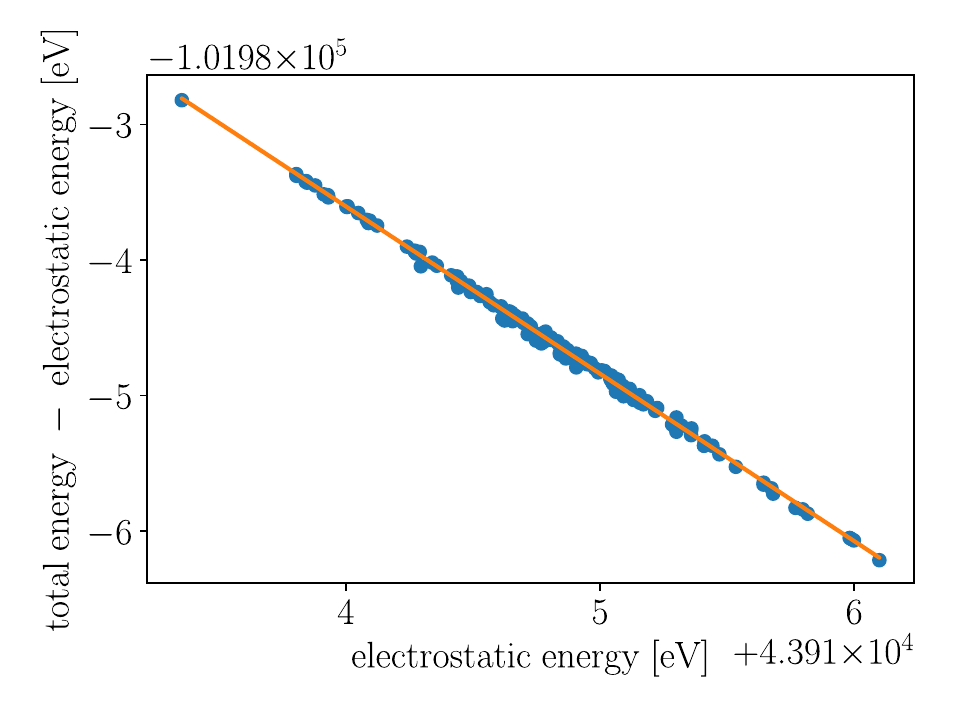}
\caption{Total energy minus classical electrostatic energy versus the
classical electrostatic energy in the ArKr mixture.}.
\label{fig:mix-correlation}
\end{figure*}
\begin{figure*}[t]
\centering
\includegraphics[width=\textwidth]{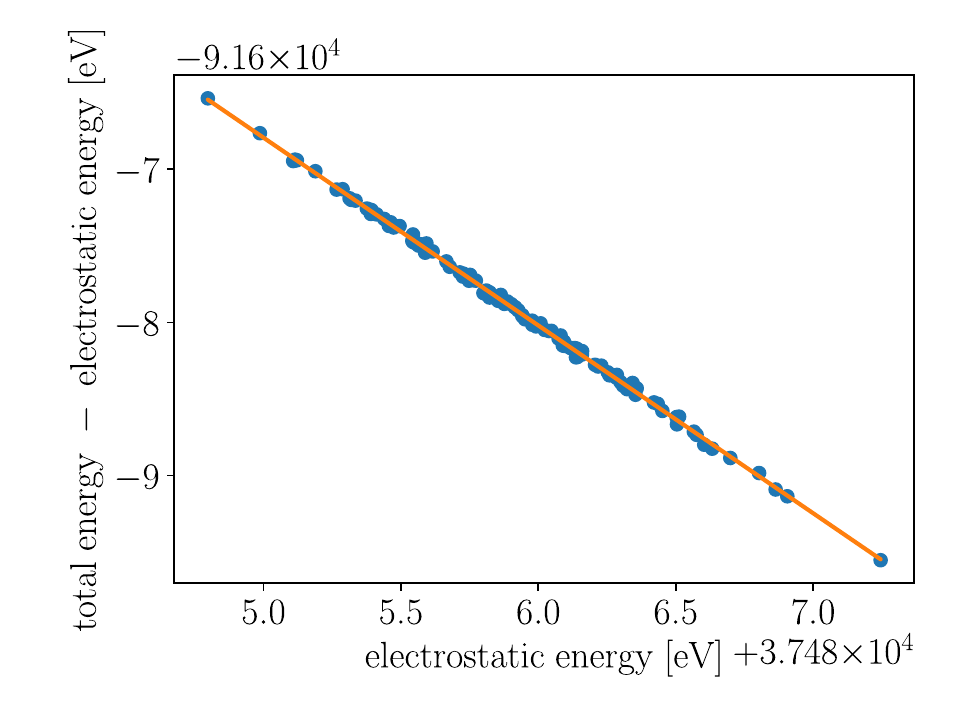}
\caption{Total energy minus classical electrostatic energy versus the
classical electrostatic energy in the pure Kr system.}.
\label{fig:kr-correlation}
\end{figure*}

\section{Appendix B: Architectures}
Here we outline the results of local energy prediction investigations on different neural network architectures.
\begin{figure*}[htbp]
    \centering
        \input{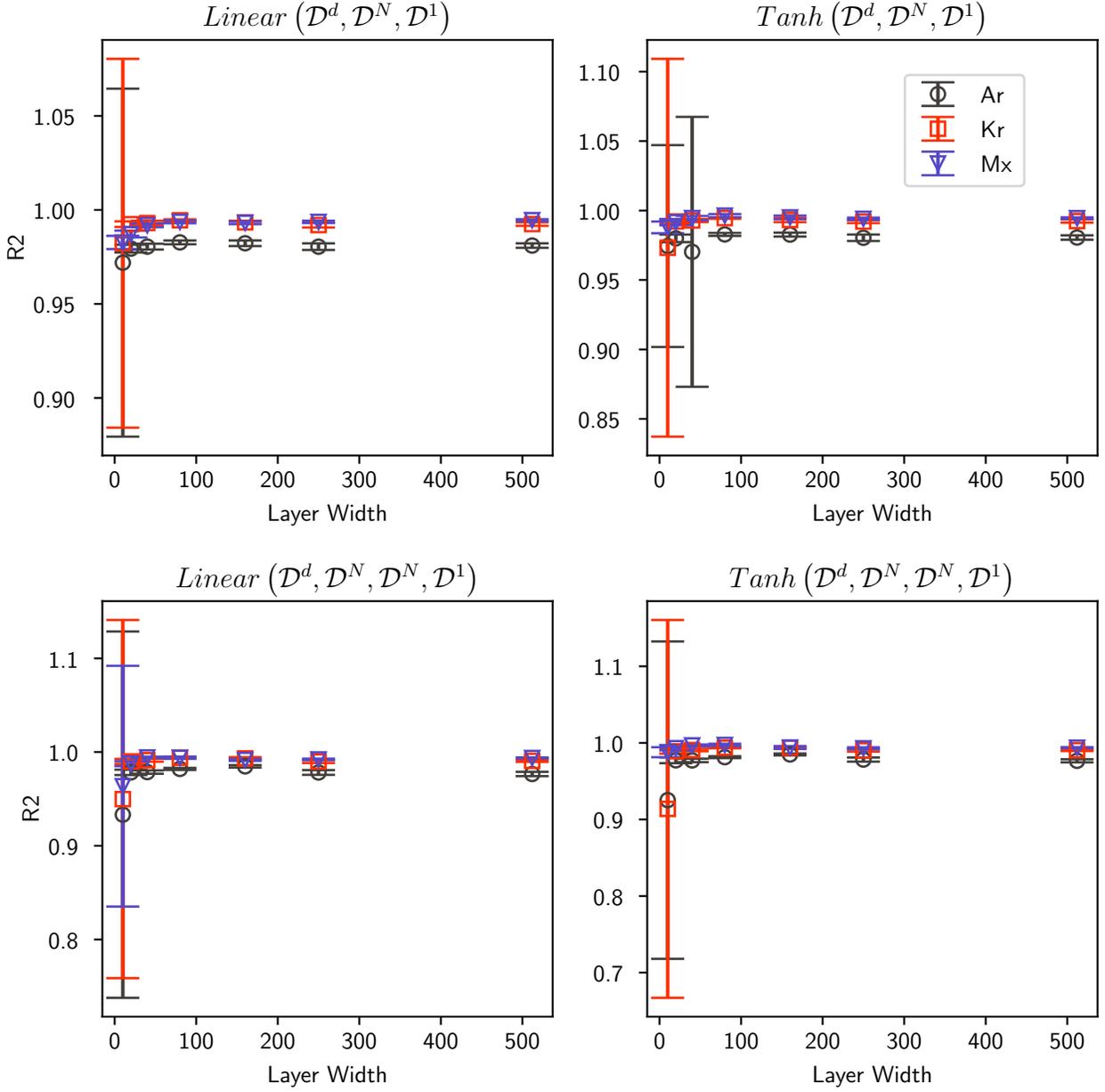}
    \caption{Correlation score for the total energy of each system. As the networks are trained on the total energy, we expect the score to remain high over the full space of architecture.}
    \label{fig:si-all-archs}
\end{figure*}

\begin{figure*}[htbp]
    \centering
        \input{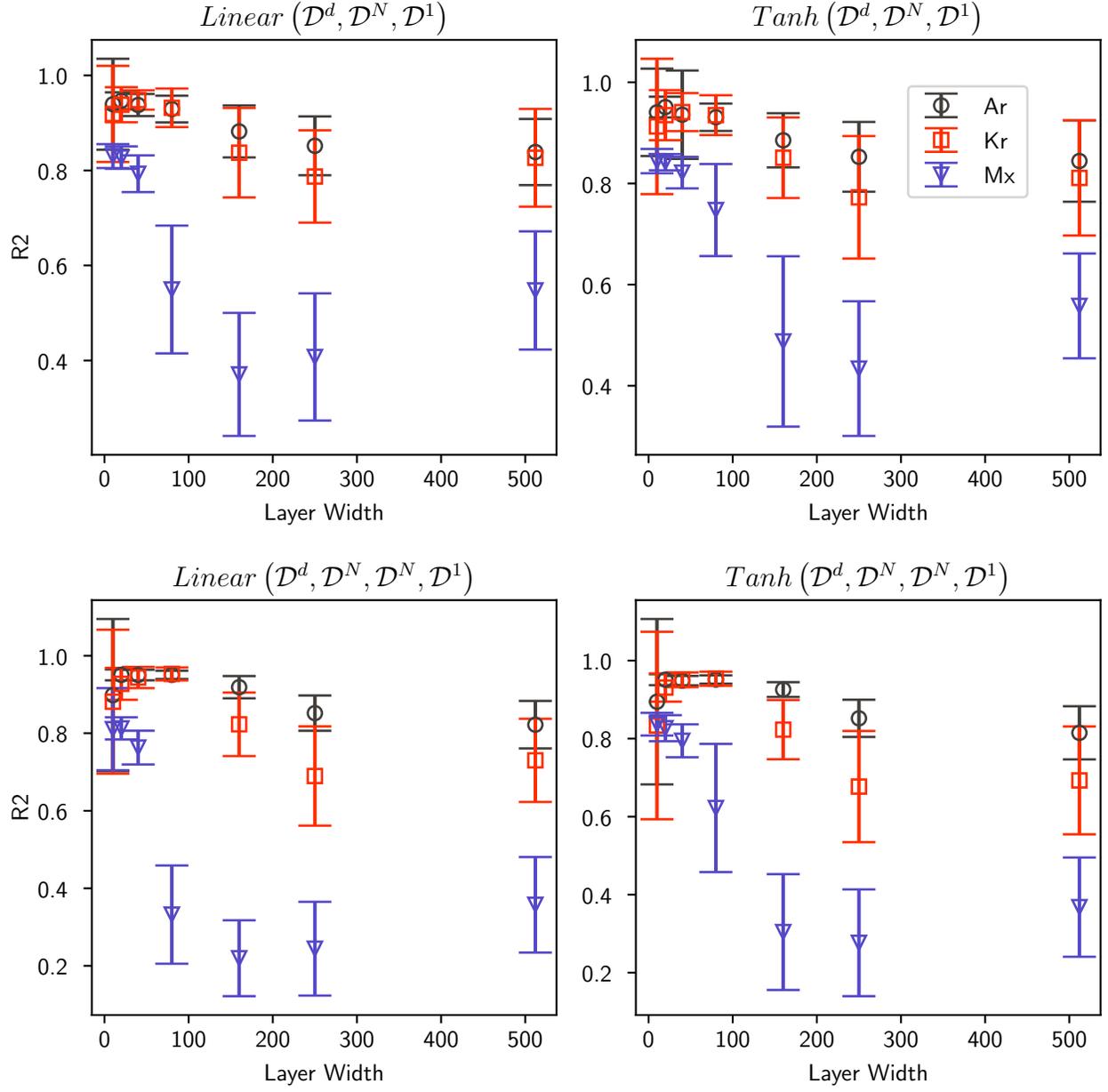}
    \caption{Correlation score for the local energy of each system.}
    \label{fig:si-local_energy}
\end{figure*}

\begin{figure*}[htbp]
    \centering
        \input{figures/figure_9.pgf}
    \caption{Correlation score for the total energy of each system computed using the transferred models.}
    \label{fig:si-total_transfer}
\end{figure*}

\begin{figure*}[htbp]
    \centering
        \input{figures/figure_10.pgf}
    \caption{Correlation score for the local energy of each system computed using the transferred models.}
    \label{fig:si-local_transfer}
\end{figure*}

\end{document}